\documentclass[aps,pre,preprint,groupedaddress,showpacs]{revtex4}

\usepackage{amsmath}

\usepackage{amssymb}

\usepackage{amsbsy}

\usepackage{graphicx}

\begin{document}

\title{Discrete soliton mobility in two-dimensional waveguide arrays with 
saturable nonlinearity}

\author{Rodrigo A. Vicencio}\email{rodrigov@mpipks-dresden.mpg.de} 
\affiliation{Max Planck Institute for the Physics of Complex Systems, 
N\"othnitzer Str. 38, D-01187 Dresden, Germany}

\author{Magnus Johansson}\email{mjn@ifm.liu.se}
\homepage{http://www.ifm.liu.se/~majoh}
\affiliation{Max Planck Institute for the Physics of Complex Systems,  
N\"othnitzer Str. 38, D-01187 Dresden, Germany}
\affiliation{Department of Physics, Chemistry and Biology (IFM), Link\"{o}ping 
University, SE-581 83 Link\"{o}ping, Sweden}
\affiliation{University of Kalmar, Department of Chemistry and Biomedical 
Science, SE-391 82 Kalmar, Sweden}

\begin{abstract}

We address the issue of mobility of localized modes in two-dimensional 
nonlinear Schr\"odinger lattices with saturable nonlinearity. This describes 
e.g.\ discrete spatial solitons in a tight-binding approximation of 
two-dimensional optical waveguide arrays made from photorefractive crystals. 
We discuss numerically obtained 
exact stationary solutions and their stability, focussing on three different 
solution families with peaks at one, two, and four neighboring sites, 
respectively. When varying the power, there is a repeated exchange of 
stability between these three solutions, with symmetry-broken families of 
connecting intermediate stationary solutions appearing at the bifurcation 
points. When the nonlinearity parameter is not too large, we observe good 
mobility, and a well defined Peierls-Nabarro barrier measuring the minimum 
energy necessary for rendering a stable stationary solution mobile. 

\end{abstract}

\date{\today}

 \maketitle

\section{Introduction}

There is a large current interest in nonlinear mechanisms for storage and 
transport of localized coherent packages of energy in spatially periodic 
systems, which may be described by discrete lattice models in one, two, or 
three spatial dimensions. Such systems quite generally may support intrinsic 
localized modes, or discrete breathers, which are exact, generally 
time-periodic, spatially localized solutions to the underlying dynamical 
lattice equations and, under certain conditions, also may show mobility. For 
a recent review of properties and applications of such intrinsic localized 
modes, see \cite{CFK04}. In particular, within nonlinear optics such localized 
modes may appear as 
discrete spatial solitons in periodic structures describing arrays of coupled 
waveguides. See, e.g., \cite{SKES03, CLS03} for reviews of experimental 
observations as well as theoretical modelling of discrete spatial solitons, 
and \cite{Chen05,Matuszewski} and references therein 
for more recent experimental results.

In such systems, the longitudinal variable along the waveguides plays the role 
of time in the dynamical lattice equations, while the arrays themselves may be 
either one-dimensional (1D) (e.g.\ \cite{Eisenberg,Fleischer1D,NOKK03}) 
or two-dimensional (2D) (e.g. \cite{Fleischer2D}) transversally. 

Traditionally \cite{CJ88,Eisenberg}, most attention has been put on waveguide 
arrays 
constructed from Kerr nonlinear media, in which case the appropriate lattice 
model derived from coupled-mode theory (tight-binding type approximation) is 
the cubic discrete nonlinear Schr\"odinger (DNLS) equation 
(see, e.g., \cite{EJ03} for a review on properties and applications of this 
model). However, more recently 
(e.g.\ \cite{Fleischer1D,NOKK03,Fleischer2D,Chen05,Matuszewski} 
and references therein) much experimental effort has been devoted to 
generating discrete solitons in photorefractive media, where the nonlinearity 
is not anymore of pure Kerr type \cite{Efremidis}. In particular, 
optically-induced lattices with focusing as well as defocusing nonlinearities 
were created e.g.\ in \cite{Fleischer1D,NOKK03,Fleischer2D}, leading to 
direct observation of many earlier predicted phenomena such as 
spatial gap solitons 
\cite{Fleischer1D} and solutions of different symmetries (odd and even) in 
1D arrays \cite{NOKK03}. In addition,   
the recent 
works \cite{Chen05,Matuszewski} have reported observation of spatial 
gap solitons and self-trapping also in 1D 
permanent waveguide arrays with photovoltaic defocusing nonlinearity, 
showing strong saturation 
at higher powers. In this case, the appropriate lattice model in the 
tight-binding limit is a 1D DNLS equation with saturable on-site nonlinearity 
(a discrete version of the Vinetskii-Kukhtarev \cite{Vinetskii} equation), 
which was introduced in \cite{Stepic04} 
and further studied in \cite{HMSK04,Khare04,Maluckov05,CE05}.
It is important to remark, that the discrete model used in 
\cite{Maluckov05} to model waveguide arrays with photovoltaic defocusing 
nonlinearity is mathematically equivalent \cite{CE05} to the model 
for focusing nonlinearity of \cite{Stepic04, HMSK04}. A recent discussion 
and experimental demonstration of this equivalence 
was given in \cite{Matuszewski}.   

One remarkable property of the 1D saturable DNLS equation, discovered in 
\cite{HMSK04}, is the boundedness, and at certain points even vanishing, of 
the so-called Peierls-Nabarro (PN) potential barrier as a function of the 
soliton power. The PN barrier for DNLS-like equations may be defined 
\cite{KC93} as the difference in energy (Hamiltonian) between the two 
fundamental localized modes centered at respectively in-between lattice sites, 
at the same power (norm) (the latter being a conserved quantity for the 
dynamics). Thus, it is expected to give a lower bound to the amount of 
additional energy necessary to render a stable stationary solution mobile. It 
was also numerically confirmed \cite{HMSK04,Maluckov05,CE05} that the mobility 
of high-power localized solutions in the saturable 1D DNLS model was 
considerably enhanced compared to the cubic DNLS model, where the PN-barrier 
grows rapidly as the power increases \cite{pnpMor}. Another peculiar property 
of the 1D saturable DNLS model is the existence of a family of exact 
analytical sech-shaped solutions, centered at arbitrary lattice positions, for 
some particular regimes of parameter values \cite{Khare04}. The power of these 
solutions, which are linearly marginally stable, depends continuously on the 
position of their center, and it can be shown that this solution family exists 
in a neighborhood of the first zero of the PN barrier, bifurcating with the 
site-centered and bond-centered solution families exactly at the points where 
these families exchange their stability. Thus, the analytical solutions 
constitute a family of `intermediate' symmetry-broken solutions connecting 
site-centered and bond-centered solutions, analogously to the scenario for 
enhanced mobility described for another type of extended 1D DNLS equation in 
\cite{OJE03}, as well as for more general chains of coupled oscillators in 
\cite{Cretegny98}.

However, for 2D arrays, much less is known about discrete soliton mobility. 
Essentially, as was described in \cite{Christiansen96}, for cubic nonlinearity 
only wide discrete solitons are mobile. However, these are unstable to a 
`quasicollapse' process, so that after moving a few lattice sites the broad 
solitons self-focus into narrow localized peaks, which get pinned by the 
lattice \cite{Christiansen96}. 
It is our purpose here, to show that a saturable nonlinearity may, under 
certain conditions, lead to highly mobile discrete solitons
also for 2D waveguide arrays. The study of discrete soliton mobility 
has been suggested to be one of the most important issues in the 
implementation of these theoretical concepts in all-optical switching schemes 
for one-dimensional nonlinear arrays \cite{rodrigo} and very recently for 
two-dimensional ones in a `reduced' geometry \cite{reduced}. For this 
reason, the two-dimensionality can be viewed as a large improvement in 
this direction, because of the promising possibility of multiple-site 
connections and the direct integration with photonic crystals.

An interesting analogy may be drawn to what is 
known within the field of polarons, where in the case of the standard 
semiclassical Holstein model with harmonic on-site oscillator potentials the 
stable 
(small) polarons are always pinned to the lattice in 2D, while if a realistic 
saturable anharmonicity is taken into account for the oscillator potentials, 
moving polarons may also exist \cite{ZCR98}. In the former case, the static 
polarons are obtained as stationary solutions to the cubic DNLS equation, 
while in the latter case as solutions to the saturable DNLS equation (although 
the dynamics of the polarons is more complex as it involves electron as well 
as phonon degrees of freedom).

The structure of this paper is as follows. In Sec.\ \ref{Model} we describe 
the 2D saturable DNLS equation, its basic properties, and the region of 
existence for localized solutions. Next, in Sec.\ \ref{Stabi} we analyze the 
stability of these solutions and we introduce the concept of Intermediate 
Solutions. In Sec.\ \ref{Mobi}, we present our results for the mobility of 
localized solutions in 2D arrays. Finally, Sec.\ \ref{Conclu} concludes the 
paper. 

\section{Model}
\label{Model}

We consider the following (general) form of the 2D saturable DNLS equation for 
an isotropic medium, analogous to the 1D model in \cite{Stepic04,HMSK04,CE05}, 
\begin{equation}
i \frac{\partial u_{n,m}} {\partial \xi} + \Delta u_{n,m} -\gamma\frac{u_{n,m}}
{\left(1+|u_{n,m}|^2\right)} = 0 ,\label{sat} 
\end{equation}
where $\xi$ is the normalized propagation distance, $u_{n,m}$ describes the 
electric field amplitude in the $\{n,m\}$ site, $\Delta$ represents the 2D 
discrete Laplacian, 
$\Delta u_{n,m} \equiv u_{n+1,m}+u_{n-1,m}+u_{n,m+1}+u_{n,m-1}$. 
The parameter $\gamma$ is given by the ratio between the nonlinear 
parameter 
and the coupling constant \cite{Stepic04,HMSK04,CE05}. 
We choose  $\gamma>0$ in our computations without loss of generality. Note 
that, although this implies that we are formally restricting to a focusing 
nonlinearity, our results are immediately translated to the defocusing case 
through the staggering transformation 
$u_{n,m} \rightarrow (-1)^{m+n} u_{n,m} , \xi \rightarrow -\xi$. 
We use an isotropic approximation which essentially considers the 
coupling between neighboring sites in the $\hat{n}$ and $\hat{m}$ 
directions as equal. In the experiment, the intrinsic anisotropy of 
photorefractive materials can be reduced by changing the lattice orientation 
relative to the crystal axis \cite{iso}. The two conserved 
quantities for (\ref{sat}) are the energy (Hamiltonian) 
\begin{eqnarray}
H =-\sum_{n,m=1}^{N,M} \left[ (u_{n+1,m}+u_{n,m+1})\ u_{n,m}^{*}\right. 
\nonumber\\ \left.-\frac{\gamma}{2} \ln (1+|u_{n,m}|^2) +c.c. \right], 
\label{H}
\end{eqnarray}
and the Power (norm)
\begin{equation}
P=\sum_{n,m=1}^{N,M} |u_{n,m}|^2. \label{P}
\end{equation}
Thus, for small $P$, Eq.\ (\ref{sat}) reduces to the cubic DNLS equation with 
focusing nonlinearity of strength $\gamma$, which then may be replaced by 
unity through rescalings, while for larger $P$ saturation effects become 
important and $\gamma$ is left as an independent parameter. Note that slightly 
different forms of the saturable DNLS equation were considered (in 1D) in 
\cite{Khare04,Maluckov05}, but these are easily shown to be equivalent to the 
model of \cite{Stepic04,HMSK04,CE05} through simple rescalings and gauge 
transformations \cite{CE05}. In \cite{Stepic04,HMSK04,Maluckov05} the authors 
study analytically the case of $\gamma\approx 9$, based on experimental 
parameters. We decided to study, for the two-dimensional array, a similar 
case for $\gamma=10$. As it further will be explained, we also study the case 
of $\gamma=4$ where we found that a good mobility can be observed for lower 
level of power ($P$).

To study the dynamics we introduce a definition of the soliton center in the 
direction $\hat{n}$ (horizontal direction) as a function of $\xi$,
\begin{equation}
<n>\left(\xi\right)\equiv \sum_{n,m=1}^{N,M} n\ |u_{n,m}(\xi)|^2/
\sum_{n,m=1}^{N,M} |u_{n,m}(\xi)|^2, \label{ave}
\end{equation}
with  an equivalent definition in the vertical direction $\hat{m}$ 
for $<m>(\xi)$. We define ``axial propagation'' as the propagation in the 
$\hat{n}$ ($k_x\neq0, k_y=0$) or in the $\hat{m}$ ($k_y\neq0, k_x=0$) 
directions, and ``diagonal propagation''  when we launch the solution with the 
same angle in both directions ($k_x=k_y\neq0$).

\begin{figure}[tbp]\centerline{\scalebox{0.6}{
\includegraphics{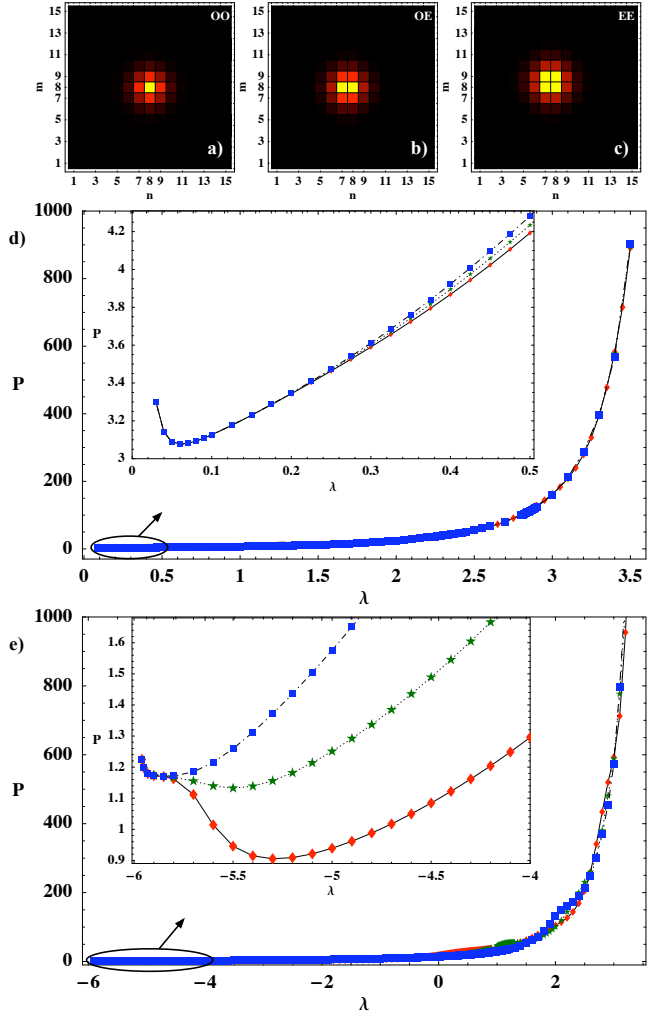}
}}
  \caption{(Color online). (a) OO profile. 
(b) OE profile. (c) EE profile. (d) and (e) $P$ versus $\lambda$ for $\gamma=4$ and for $\gamma=10$, respectively. 
Diamonds, stars, and squares represent OO, OE, and EE solutions, respectively.
System size $N=M=15$ ((d) and (e) insets: $N=M=31$).}
  \label{fig0}
\end{figure}

Stationary solutions to (\ref{sat}) are those of the form 
$u_{n,m}(\xi)=u_{n,m} e^{i \lambda \xi}$, where $\lambda$ represents the 
spatial frequency in the propagation direction. As we will mainly consider the 
fundamental  
localized solutions, we will assume $u_{n,m}$ to be real (although  
stationary localized solutions with nontrivial complex $u_{n,m}$, 
such as vortex solitons, generally also exist
in 2D photorefractive optical lattices \cite{Yang}). Using a 
Newton-Raphson 
method, we have found the same types of stationary solutions as described, 
e.g., in Ref.\ \cite{2Dkevre} for the 2D cubic model. Using the usual optical 
terminology \cite{Eisenberg}, we rename these 
solutions: one-peak solutions, as Odd-Odd (OO) (Fig.\ \ref{fig0}(a)); 
two-peaks solutions, as Odd-Even (OE) (Fig.\ \ref{fig0}(b)); 
four-peaks solutions, as Even-Even (EE) (Fig.\ \ref{fig0}(c)). 
%
%%%%%%FIG0
The region of existence (see Fig.\ \ref{fig0}) of localized solutions in this 
system has a very different structure compared to the cubic DNLS case 
\cite{Mez,2Dkevre}. 
This can be understood by considering the properties of plane wave-solutions 
to 
Eq.\ (\ref{sat}) in the limits of small and large amplitudes, respectively. 
For small-amplitude plane waves, it is easy to 
show, by neglecting the term  $|u_{n,m}|^2$ in the denominator of the 
last term in 
Eq.\ (\ref{sat}), 
that the linear band corresponds to 
$\{-\gamma-4,-\gamma+4\}$; the superior limit ($\lambda=0$ for $\gamma=4$ and 
$\lambda=-6$ for $\gamma=10$, see Fig.\ \ref{fig0}) corresponding to 
constant-amplitude solutions (zero wave vector) and being the border where 
small-amplitude 
delocalized and localized solutions are connected. On the other hand, in the 
high-amplitude limit, we 
may completely neglect the last term in Eq.\ (\ref{sat}) 
due to the saturable nature of the nonlinearity, and thus in this limit 
plane waves constitute the band $-4\leq \lambda \leq 4$ (independent of 
$\gamma$); again with the superior limit $\lambda=4$ corresponding to 
the constant-amplitude (zero wave vector) solution.  
The superior limit for 
localized solutions is observed to be $\lambda=4$ by increasing the frequency 
(see Fig.\ \ref{fig0}), and thus 
the region of existence for localized modes is between the low-amplitude 
and high-amplitude limits for the upper band edge (zero wave vector) plane 
wave. 

Note that, similarly to the cubic 2D DNLS equation 
\cite{Mez,Christiansen96,Flach,2Dkevre}, the power for the 
localized solutions is non-zero in the small-amplitude limit, and, when 
the frequency increases, the power first decreases until it reaches a minimum. 
When the frequency of the localized solutions increases further, the power 
increases indefinitely. However, the nonlinearity is saturable, 
which means that there is some threshold value for the amplitudes beyond which 
localization effects should diminish. For 
instance, if we take an OO solution (Fig.\ \ref{fig0}(a)), 
we can see that when 
the central site achieves such an amplitude threshold value, the surrounding 
sites begin to increase their amplitudes, 
delocalizing the profile and evolving to 
a high-amplitude plane wave. An analogous scenario was discussed, in a polaron 
context, in Ref.\ \cite{ZCR98}, and similar properties can also be described 
for the 1D saturable problem \cite{Stepic04,HMSK04,Maluckov05}.

\section{Stability analysis of two-dimensional localized solutions}
\label{Stabi}

Linear stability of stationary solutions may be investigated 
in a standard way (see, e.g., Ref.\ \cite{CE85}) 
by writing $u_{n,m}(\xi)=(u_{n,m}+\phi_{n,m}(\xi)) e^{i \lambda \xi}$, 
leading to the linearized equation
for $\phi_{n,m}$, the perturbation function (PF). To solve this problem 
numerically, we use the technique outlined in Ref.\ \cite{Khare04} for the 
stability analysis of localized solutions in a similar nonlinear discrete 
medium. We split the PF into real and imaginary parts, 
$\phi_{n,m}=x_{n,m}+i y_{n,m}$ ($x, y \in \Re$), and insert them into the 
linearized equation.  
Taking a two-dimensional square discrete array ($N=M$) to compute stationary 
solutions and to perform their stability analysis, 
we proceed to map the 2D problem to a 1D representation. 
Defining $\overrightarrow{Z}$ as the vector ${Z_1,...,Z_{N^2}}$ ($Z=X + iY$), 
we can write the equations for the perturbation functions in a one-dimensional 
representation as
\begin{eqnarray}
\dot{\overrightarrow{X}}+\mathbf{A} \overrightarrow{Y}=0\ \text{and}\
\dot{\overrightarrow{Y}}-\mathbf{B} \overrightarrow{X}=0 \Rightarrow 
\nonumber\\
\ddot{\overrightarrow{X}}+\mathbf{A B} \overrightarrow{X}=0 \ 
\text{and}\ \ddot{\overrightarrow{Y}}+\mathbf{B A} \overrightarrow{Y}=0,
\end{eqnarray}
where $\mathbf{A}$ and $\mathbf{B}$ are $N^2\times N^2$ matrices, depending of 
the $u_{n,m}$ profiles.
Now, the stability analysis consists in finding the eigenvalues of the 
matrix $\mathbf{A B}$ (the matrix $\mathbf{B A}$ has the same eigenvalues 
\cite{Khare04}). If all the eigenvalues are 
positive the solutions of the problem 
(PF's) are oscillatory functions, which implies stability. On the other hand, 
a negative or complex 
eigenvalue means the existence of exponentially increasing PF's, implying 
instability. 

For a given frequency $\lambda$ we compute the Power $P$, the 
Hamiltonian $H$, and the $\mathbf{A B}$ eigenvalues for the OO, OE, and EE 
solutions. 
%
%%%%%%FIG1
\begin{figure}[tbp]\centerline{\scalebox{0.6}{
\includegraphics{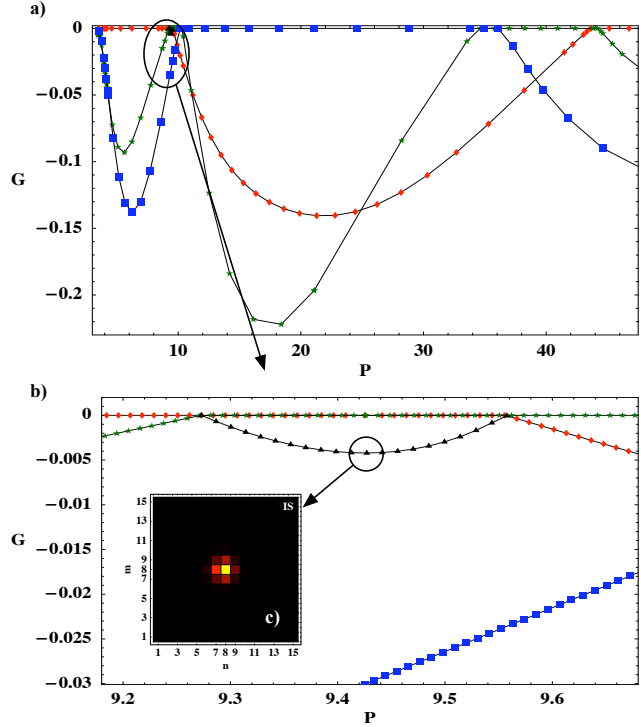}
}}
  \caption{(Color online). Results for $\gamma=4$. (a) Smallest eigenvalue $G$ versus Power. 
(b) Zoom of (a) in the 
region $P\sim 9.4$. (c) Profile of intermediate solution (IS). 
Diamonds, stars, squares, and triangles 
represent OO, OE, EE, and IS solutions, respectively.}
  \label{fig1}
\end{figure}

\begin{figure}[tbp]\centerline{\scalebox{0.6}{
\includegraphics{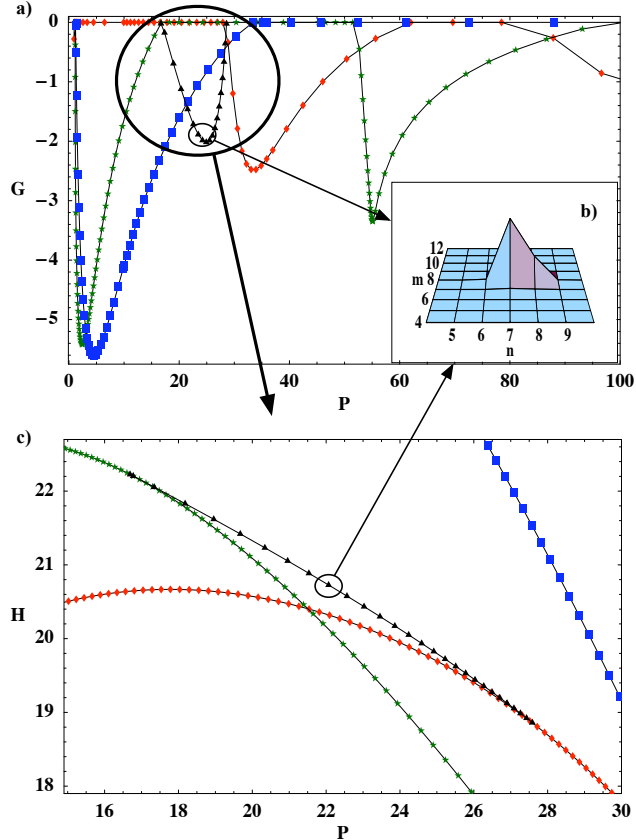}
}}
  \caption{(Color online). Results for $\gamma=10$. (a) Smallest eigenvalue $G$ 
versus Power. (b) Profile of intermediate solution (IS). (c) 
Energy $H$ versus Power for the region enclosed with a circle in (a). 
Diamonds, stars, squares, and triangles represent OO, OE, EE, and IS 
solutions, respectively.}
  \label{fig2}
\end{figure}

We take the most negative eigenvalue of each solution, calling it ``$G$'', 
and we plot it for different powers in Fig.\ \ref{fig1}(a) and (b) for 
$\gamma=4$, and in Fig.\ \ref{fig2}(a) for $\gamma=10$. 
$G=0$ or $G<0$ implies stability or instability, respectively (there are no 
complex eigenvalues for these solutions). 
An oscillatory behavior in the stability of solutions can clearly be viewed 
in these figures. Depending on the level of Power, a change in stability for 
different solutions may be observed. This result is in direct concordance 
with the previous result showed in Ref.\ \cite{HMSK04} for the 
one-dimensional problem. However, the two-dimensional problem is richer in 
properties since it has three main localized solutions, and because, as we 
will show below, the 
exchange of stability produces good mobility of very localized solutions for 
2D arrays. 

One way to understand the concept of exchange of stability between two 
different solutions is to look for Intermediate Solutions (IS), as was done 
for 1D lattices in \cite{Cretegny98, OJE03}. Such solutions, having 
symmetry-broken profiles 
interpolating between the two solutions which exchange their stability,
typically only 
exist in a limited parameter regime, connecting through bifurcations with 
the two solutions of higher symmetry.
Generally, as discussed in \cite{OJE03}, the IS may be either linearly stable, 
in which case the two symmetric 
solutions are simultaneously unstable, or the IS may be 
unstable, when the two symmetric 
solutions are simultaneously stable. 
Here, we find IS connecting two solutions that share stability, 
and thus these IS  are unstable.

For $\gamma=4$, we found that the first ``exchange region'' 
(defined as the region where two solutions are stable simultaneously) 
is between the OO and the OE solutions. 
This region can be observed in  Fig.\ \ref{fig1}(b), 
where we also present the profile for the IS (Fig.\ \ref{fig1}(c)). 
For this value of $\gamma$, we can see in Fig.\ \ref{fig1}(a) that the 
exchange regions when increasing the power are, consecutively: 
OO-OE, OE-EE, EE-OE, OE-OO. 
Multiple crossing points between the three different solutions can be 
observed by plotting the Hamiltonian versus Power, coinciding with the 
exchange regions defined before. These energy crossings imply that the new 
stable solution has lower energy compared to the others. This can be 
interpreted as an oscillation in the PN barrier \cite{HMSK04}. However, there 
is no formal definition of this barrier for 2D arrays, and it generally 
must depend on the direction of motion. We may define it loosely 
as the largest 
difference 
in energy, at constant norm, between two stationary solutions of the system, 
close to which a localized mode must pass when moving adiabatically through 
the lattice in a certain direction. 
(Note that, even in 1D, the simple definition of the PN barrier 
used e.g.\ in Ref.\ \cite{HMSK04}
as the difference between site-centered and bond-centered solutions is 
generally
not appropriate when these are simultaneously stable, since the energy of the 
IS may be considerably larger). 
The behavior of this barrier in arrays with saturable nonlinearity is 
remarkable. As seen from Fig.\ \ref{fig1}(a), for each of the three 
fundamental localized solutions, we can 
pass from a 
region where the solution is unstable to one where it is stable, by tuning 
the corresponding solution power (analogously to the 1D scenario shown in 
Ref.\ \cite{HMSK04}). As we show explicitly in Sec.\ \ref{Mobi}, this exchange 
of stability properties for different solutions implies an improvement in the 
mobility for discrete arrays, which is an important aspect for the 
implementation of this concept in all-optical switching systems. 
(We do not here explicitly show the results for 
$H$ versus $P$ for $\gamma=4$, since the 
relative differences between the respective energies are too small to be 
clearly represented graphically. See the discussion in Sec.\ \ref{Mobi} for
some typical values.) 
%
%%%%%%FIG2

Compared to the previous case, for $\gamma=10$ (implying a higher 
nonlinearity or a lower coupling between sites) solutions are more localized 
as expected. The stability diagram is presented in Fig.\ \ref{fig2}(a). As in 
the former case, multiple exchange regions where two solutions are stable at 
the same level of power can be observed: OO-OE, OE-EE, EE-OO, EE-OE. It is 
very clear from Fig.\ \ref{fig2}(a) that the IS (Fig.\ \ref{fig2}(b)) exists 
in the exchange region where the OO and OE solutions are stable 
(region enclosed 
by a circle in Fig.\ \ref{fig2}(a)). 
In this case the differences in energies are 
bigger, and we show in Fig.\ \ref{fig2}(c) a plot for the Hamiltonian versus 
Power. In this figure the way in which the IS connect the OE and the OO 
solutions can be very well observed. Before the crossing point, the energy 
difference $\Delta H\equiv H_{\mathrm {OO}}-H_{\mathrm {OE}}$ is lower than 
zero. After the crossing point this difference has changed its sign. This is 
a clear confirmation that the oscillation in the PN potential, if defined 
in the ``naive'' way as $\Delta H$, is due to the 
exchange in the stability of different solutions.
The OE solution becomes stable (see 
Fig.\ \ref{fig2}(a)), then it shares stability with the OO solution, and 
finally it continues to be stable while the OO solution becomes unstable. 
The same behavior is described by energy considerations from 
Fig.\ \ref{fig2}(c).  Note however, that the 
actual energy barrier to overcome for a localized solution moving 
adiabatically 
in an axial 
direction is larger than $\Delta H$ in the regime of simultaneous OO and OE 
stability, if the path goes via the IS whose energy 
is larger. This will be illustrated in Sec.\ \ref{Mobi}.

\section{Mobility of two-dimensional localized solutions}
\label{Mobi}

We study the mobility of 2D discrete solitons by solving numerically 
Eq.\ (\ref{sat}) for initial conditions obtained by slightly perturbing 
the stationary solutions described above. 
To be specific, we here choose the OO 
solution and study mobility in regions where this profile is always stable. 
In our simulations, we always consider perturbations obtained by  
``kicking'' the initial OO solutions using: 
$u_{n,m}(0)=u_{n,m} \exp[i \left(k_x n +k_y m\right)]$. 
Note that such perturbations do not change the power, but generally increase 
the energy compared to the stationary solutions.

\begin{figure}[htb]\centerline{\scalebox{0.6}{
\includegraphics{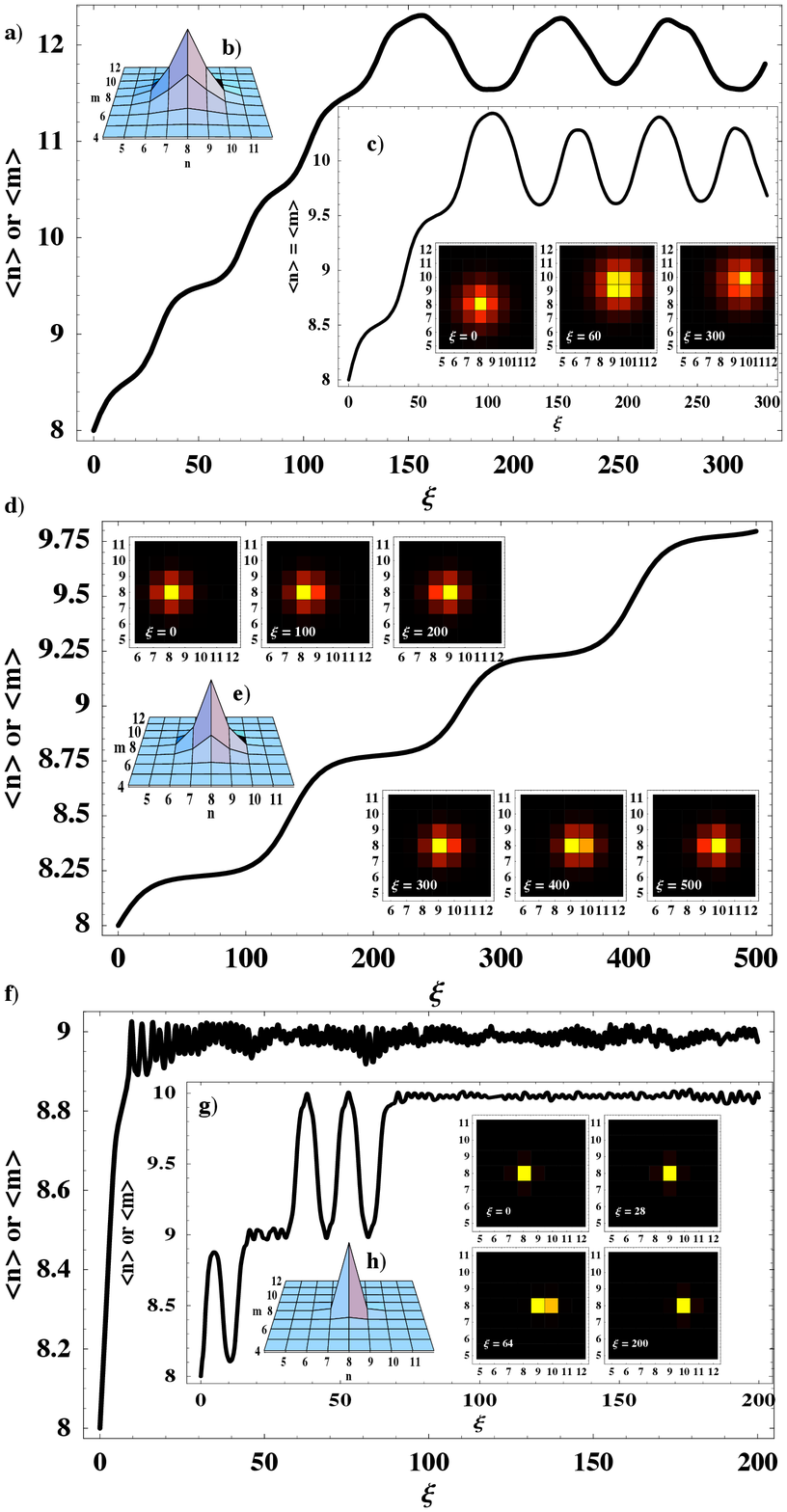}
}}
  \caption{(Color online). Mobility results. $\gamma=4,\ P=3.99$: (a) 
axial propagation for $k_x$ or $k_y=0.0323$; (b) OO profile; 
(c) diagonal propagation for $k_x=k_y=0.031$. $\gamma=4,\ P=9.42$: 
(d) axial propagation for $k_x$ or $k_y=0.00624$; (e) OO profile. 
$\gamma=10$: (f) axial propagation for $k_x$ or $k_y=0.6$ ($P=20.43$); 
(g) axial propagation for $k_x$ or $k_y=0.5$ ($P=24.76$); (h) OO profile. 
System size $N=M=15$, periodic boundary conditions. }
  \label{fig3}
\end{figure}

In Fig.\ \ref{fig3} we show dynamical results for $\gamma=4$ and 
$\gamma=10$. First, for $\gamma=4$, we study the dynamics for small-power 
solutions, $P\approx4$. Fig.\ \ref{fig3}(a) shows the average axial position 
(\ref{ave}) for an OO profile (belonging to the branch with 
$\partial P / \partial \lambda >0$ in Fig.\ \ref{fig0}(d)) 
kicked in the axial direction. In this region of 
power the OO mode is stable, while the OE and EE modes are both 
unstable (see Fig.\ \ref{fig1}(a)). The additional energy due to 
the kick corresponds to $\Delta H=0.0038$, which is larger than the 
energy difference between the OO and OE configurations, 
$\Delta H=0.0032$. For this reason, the solution first begins to move, 
but is then
trapped 4 sites away from the input center in a (stable) OO configuration. 
Due to radiation losses, it has not 
sufficient energy to overcome the next barrier, but instead begins to 
oscillate around its new center due to the excitation of its stable internal 
mode.
An interesting detail in this result is the structure of the curve. 
When the OO mode propagates, it changes its form between the OO and OE 
profiles. It is clear from the figure that the maximum propagation velocity 
corresponds to the OO configuration (bigger slope, potential well), and the 
minimum velocity (slope $\approx0$, saddle point) corresponds to the OE 
configuration (centered between two sites in the propagation direction). 
Similar results for 1D systems have been described, e.g., in Refs.\  
\cite{Cretegny98, GJ03}.
%
%%%%%%FIG3

For the same parameter values, we illustrate in Fig.\ \ref{fig3}(c) 
the propagation of an OO profile kicked in the diagonal direction. 
This plot describes the identical evolution in 
both axes. It can be observed that the OO profile propagates in the diagonal 
direction, changing its form from an OO configuration to an EE one (see the 
inset). The final switching here was two sites in both directions, 
from the input position $\{8,8\}$ 
to $\{10,10\}$. Here, the maximum propagation velocities correspond to the OO 
configurations, and the almost zero velocities correspond to the EE 
configurations. As before, the soliton will not 
be able to continue its propagation because of the radiation losses. 
In this case, the diagonal kick corresponds to 
$\Delta H =0.007$, while the energy difference between OO and EE 
configurations at this power is $\Delta H =0.006$. 

The dynamics shown in 
Figs.\ \ref{fig3}\ (a) and (c) corresponds essentially to that of a typical 1D 
cubic DNLS case in the regime of low power, where the PN barrier is relatively 
small, and the site-centered solution possesses a symmetry-breaking internal 
translational mode (see, e.g., \cite{JAGCR98}), 
which may be excited by kicking the solution.
Now, if we compare with the 2D cubic DNLS case, the dynamics and the 
regions of existence of the stationary localized solutions are very 
different. 
First, in a P 
versus $\lambda$ diagram for the 2D cubic DNLS case \cite{2Dkevre}, there are 
different power thresholds for OO, OE, and EE solutions, far away separated 
in power. This implies that below a certain level of the power,
it is not possible to move the OO
profile since no OE or EE solution exists at the same power. 
If we consider a higher 
level of the power (on the high-frequency branch with a stable OO solution), 
where two or three solutions exist, the profiles are too 
localized and do not possess any localized symmetry-breaking internal 
modes. This implies also that the PN potential is large, and mobility is not 
possible by kicking the solutions since there are no internal translational 
`depinning' modes to excite. Thus, for cubic 2D DNLS, mobility is possible 
only on the `continuumlike' low-frequency branch with 
$\partial P / \partial \lambda < 0$, where OO, OE, and EE solutions are all
unstable, 
but such moving solutions will typically `quasicollapse' into localized pinned 
solutions as described in \cite{Christiansen96}.

In the 2D saturable case, these 
thresholds still exist but are much closer in power. For $\gamma=4$, the 
power thresholds for different solutions are: 
$P_{\mathrm {OO}}\approx P_{\mathrm {OE}}\approx P_{\mathrm {EE}} \approx3.07$ 
(cf.\ Fig.\ \ref{fig0}(d)). 
This means that all 
solutions have essentially the same power threshold value, and that we are in 
principle able to observe good mobility in the low-power regime of 
the branch with $\partial P / \partial \lambda > 0$ for
$P\gtrsim 3.08$. For $\gamma=10$, the differences in power are higher than in 
the 
previous case, but they are still rather small. 
The corresponding threshold values 
are (see Fig.\ \ref{fig0}(e)): 
$P_{\mathrm{OO}}= 0.90$, $P_{\mathrm{OE}}= 1.13$, 
and $P_{\mathrm{EE}}= 1.17$.
This behavior is another remarkable property of the saturable 
nonlinearity. As is wellknown (see, e.g., \cite{Flach}),  
in two-dimensional 
discrete nonlinear systems with an effectively cubic (or stronger) 
nonlinearity, a power threshold value 
always exists for localized solutions. 
Our results suggest that, increasing the value of $\gamma$, 
it is  possible to decrease this power threshold value towards zero 
(for $\gamma=100$ this value is $P_{\mathrm{OO}}=0.059$). In the 
small-amplitude regime, the dynamics is essentially governed by the cubic 
term in the Taylor expansion of the saturable nonlinearity, and thus 
Eq.\ (\ref{sat}) is well approximated by the cubic DNLS model with effective
nonlinearity parameter $\gamma P$. Thus, the threshold power should scale as 
$\gamma ^{-1}$ for large $\gamma$. A similar result was also discussed 
in Ref.\ \cite{ZCR98} (see, e.g., Fig.\ 6 in this paper).

Now, we go further to study the exchange regions where we observe 
``multistability'' of solutions. In the region of power shown in 
Fig.\ \ref{fig1}(b), we study the first exchange region between the OO and 
the OE solutions. In Fig.\ \ref{fig3}(d) we show the effect of kicking the OO 
solution in the axial direction. The energy added to the profile due to the 
kick is $\Delta H =0.0003$, and the energy difference between the OO and 
the OE configurations is $\Delta H = -0.0001$, i.e., the OE 
solution has the lowest energy. 
However, in this case there exists also an intermediate solution 
(Fig.\ \ref{fig1}(c)), which is important to explain the mobility we 
observe in this dynamics. 
The energy difference between the OO and the IS solutions is $\Delta H=0.0002$.
In fact, the solution can move 
across the array as far as the radiation losses permits it. The regions in 
which the profile changes its velocity can clearly be observed in this 
figure. First, for the OO and the OE solutions the velocity has maxima, 
which is corresponding with the stability analysis where both solutions are 
stable. Both solutions correspond to a potential well in a dynamical 
representation. We can also note, that the velocity is larger at 
half-integer values than at integer values, consistent with the fact that 
the OE solution has lower energy than OO.
The minimum velocities (slope $\approx0$) clearly correspond 
to the IS between the OO and the OE profiles \cite{IS}. For this case IS have 
shown to increase mobility, essentially because all solutions are very close 
in energy and just a little kick, corresponding to approximately the 
energy difference between the OO and IS, is required to move them. An 
analogous scenario would be seen by using the kicked stable 
OE solution as initial 
condition.

Finally, we study the case for $\gamma=10$. We first 
look for mobility in the same 
region of power as in the previous case ($P\approx4$ and $P\approx9$). For 
these powers we were not able to find mobility of localized solutions, 
essentially because the stationary 
solutions are distant in energy values. 

Then, 
we study the first crossing point between the energy of 
the OO and the OE solutions 
(Fig.\ \ref{fig2}(c)). From the previous discussion, it is clear that in the 
exchange regions, the mobility of solutions depends on the IS. So, if we want 
to switch the profile in the lattice, we first have to overcome the energy 
barrier between the OO and the IS. We study two cases where we found some 
mobility. First, we take an OO profile for a power lower than the crossing 
point power ($P\approx21.5$), i.e. $P_{\mathrm{OO}}=20.43$. For this power 
both 
solutions are stable, but the OO solution has lower energy than the OE 
solution. Therefore, we expect it to be most probable to have an OO stable 
configuration at the end 
of the process by switching the profile. However, if radiation losses are 
large, the solution might in principle also be trapped in a metastable 
OE configuration, since it might not have enough energy to overcome 
the barrier between OE and OO created by IS. 
In Fig.\ \ref{fig3}(f), we show an example of switching of 
an OO mode with one site along the axial direction. The energy due to the kick 
was $\Delta H =2.147$, the energy difference between the OO and the OE 
configurations was $\Delta H =0.370$, and the energy difference between 
the OO and the IS was $\Delta H =0.672$. The figure shows how the OO 
solution starts to move very fast (large slope), 
but it is not able to continue 
its movement in the array because of the high radiation losses of the 
switching process. This is a remarkable example of discrete switching in 
nonlinear lattices. 

Now, we consider the situation where the OO power 
($P_{\mathrm{OO}}=24.76$) is higher than the crossing point power. Here, 
both OO and 
OE solutions are stable, but the OE solution has lower energy, so from 
energetic arguments we would 
expect 
the most likely final profile to be an OE mode 
(however, depending on the amount of radiation 
losses, either an OE or an OO profile may be observed as final state as 
discussed above). The dynamics resulting from a kick 
of the OO mode producing a $\Delta H =2.045$ is illustrated in 
Fig.\ \ref{fig3}(g). 
The corresponding energy difference between the OO and the 
OE modes is $\Delta H =-1.120$, and between the OO mode and the IS, 
$\Delta H =0.118$. It is clear from these numbers that we need to supply 
the OO mode with some extra energy to move it across the array even though
the energy for the OE solution is lower. This property is due to the 
oscillating behavior of the Hamiltonian. Without such crossing points,  
2D mobility for high power solutions can generally not be expected 
because of the large PN barrier. 
Fig.\ \ref{fig3}(g) shows the switching of an OO solution with two 
sites in the axial direction 
from the input position. First, the OO solution has sufficient 
energy (big slope) to overcome the IS energy barrier. Then, its velocity 
increases as it passes the OE position, and then again it slows down when 
approaching the next IS (with central position close to site 9). Now, 
it does not have enough energy to overcome this barrier, and instead it 
makes one full oscillation around the OE position, limited by the two 
unstable IS. However, during this oscillation it has recovered some of its 
energy, and can now pass the IS barrier to the position of the 
next site, where it gets temporarily trapped into an OO configuration. 
The radiation losses are still 
not too high, and the OO mode being energetically unstable decays after a 
short distance $\xi$ into a state of large-amplitude oscillations around 
the energetically stable configuration, the OE mode 
centered in $\{9.5,8\}$ (see inset in Fig.\ \ref{fig3}(g)). 
Now the solution has 
enough energy to oscillate between two sites (9 and 10). This oscillation 
produces more radiation losses, and the final profile is an OO mode centered 
in the site number 10, trapped by the barrier created by its nearby IS. 
Figs.\ \ref{fig3}(b), (e), and (h) show different profiles 
in the different regions of mobility. It is clear from these figures that the 
mobility is enhanced for less localized solutions as expected from the PN 
potential concept, but it is not forbidden for highly localized solutions as 
we have shown.

We have checked these dynamical cases also for a bigger array, where $N=M=21$. 
The quantitative results shown in Fig.\ \ref{fig3} change somewhat 
due to the extension of the 
system, but the qualitative picture for the soliton's mobility is preserved. 
We set in our computations periodic boundary conditions, which implies that 
some radiation comes back perturbing the final state of the profile. 
Therefore, the results are dependent of the boundaries. However, the important 
issue we want to emphasize here is the mobility of highly 2D localized states, 
which is independent of the array size for saturable nonlinear media.

\section{Conclusions}
\label{Conclu}

In conclusion, we have analyzed in some detail the mechanisms leading to 
mobility of localized modes in a two-dimensional DNLS-type lattice with 
saturable nonlinearity. From a practical point of view, the most important 
result from our study is the drastic enhancement of the mobility 
resulting from the saturable nature of the nonlinearity, both for low-power 
and high-power excitations. This effect, which should be observable for 
two-dimensional waveguide arrays, is in our 
opinion more remarkable than the similar effect previously reported in 1D 
\cite{HMSK04}, since stable localized excitations for pure Kerr nonlinear 
media are known to be essentially immobile in 2D. Thus, the saturability 
of the nonlinearity introduces new possibilities for power controlled 
steering and switching also for 2D arrays. As we have shown, mobility is 
not restricted to axial directions, but also steering in diagonal directions 
is possible. 

In principle, the choice of a smaller value of $\gamma$, for instance 
$\gamma=3$, could also be good for 2D mobility, because it is expected that 
the first stability 
exchange region occurs for lower level of power. However, it is 
important to notice that a decrease in the value of $\gamma$ 
(see Fig. \ref{fig0}) also implies an increasing value for the power 
threshold of the solutions, due to a balance between the nonlinearity and the 
coupling terms. We confirmed this behaviour for $\gamma=3$, where the 
power threshold was found to be $P\approx 4.84$, while the first 
stability exchange region was found to occur at $P\approx 6.98$. 
On the contrary, if we increase further the value of 
$\gamma$, for instance $\gamma=20$, we expect that the powers for the first 
exchange region will be much higher and, in this sense, it is not really 
interesting from the optical application point of view, which always requires 
low level of power.

From a more fundamental point of view, our results also give a deeper 
understanding for the mechanisms for mobility of localized modes. In 
particular, this  concerns the relation between regimes of exchange of 
stability 
between site-centered and bond-centered stationary solutions, and 
points of vanishing of 
a so-called Peierls-Nabarro (PN) barrier defined naively as the difference 
between such solutions at constant norm. As we have shown, this definition is 
not appropriate in regimes where both these solutions are simultaneously 
stable, due to the existence of unstable intermediate solutions (IS) of higher 
energy. However, redefining the PN barrier as the energy difference to 
the relevant 
IS gave good agreement with the numerically observed additional energy 
necessary for making a stationary solution mobile. An analogous scenario 
should 
exist in 1D, and gives an intuitive explanation to why, in spite of the 
repeated vanishing of the ``PN potential'' reported in Ref.\ \cite{HMSK04} 
at several 
critical powers, the mobility is good only around the first of these 
\cite{Maluckov}. At the other (larger) critical powers, the barriers created 
by the IS are expected to be very large, and no smooth path 
in phase space passing simultaneously close to all three stationary solutions 
is likely to exist.

It is also interesting to mention, that although, in 1D, both the existence of 
a particular class of IS (which were even obtained analytically) 
\cite{Khare04}, and the ``PN barrier vanishing'' \cite{HMSK04} were previously 
known for the saturable potentials, the connection between these results, and 
their relation to exchange of stability, seems so far to have gone unnoticed.

\begin{acknowledgments}

We thank Sergej Flach for many illuminating discussions on mobility in 
discrete systems. R.V.\ thanks Milutin Stepi\'{c} for useful discussions. 
M.J.\ also thanks Aleksandra Maluckov and Michael \"Oster for discussions, and 
the MPIPKS, Dresden, for hospitality during his visit. We are grateful 
to Andrey Gorbach for a careful reading of the manuscript. M.J.\ acknowledges 
partial funding from the Swedish Research Council.

\end{acknowledgments}


\begin{thebibliography}{99}
\bibitem{IS}As a method to find numerically 
the exact IS, we take the dynamical profile 
for a 
certain $\xi$ (e.g., $\xi=80$ for Fig.\ \ref{fig3}(d)) of a kicked OO 
solution, and use it as a trial solution for the Newton-Raphson scheme 
to find stationary solutions for constructing the IS spectra in 
Figs.\ \ref{fig1} and \ref{fig2}.
\end{thebibliography}
\end{document}